# Punctuated evolution of influenza virus neuraminidase (A/H1N1)

# under migration and vaccination pressures

J. C. Phillips

Dept. of Physics and Astronomy, Rutgers University, Piscataway, N. J., 08854

1-908-273-8218      jcphillips8@comcast.net

**Abstract**

Influenza virus contains two highly variable envelope glycoproteins, hemagglutinin (HA) and neuraminidase (NA). The structure and properties of HA, which is responsible for binding the virus to the cell that is being infected, change significantly when the virus is transmitted from avian or swine species to humans.  Here we focus on much smaller human individual evolutionary amino acid mutational changes in NA, which cleaves sialic acid groups and is required for influenza virus replication.  We show that very small amino acid changes can be monitored very accurately across many Uniprot and NCBI strains using hydropathicity scales to quantify the roughness of water film packages.  Quantitative sequential analysis is most effective with the differential hydropathicity scale based on protein self-organized criticality (SOC).  NA exhibits punctuated evolution at the molecular scale, millions of times smaller than the more familiar species scale, and thousands of times smaller than the genomic scale. Our analysis shows that large-scale vaccination programs have been responsible for a very large convergent reduction in influenza severity in the last century, a reduction which is hidden from short-term studies of vaccine effectiveness. Hydropathic analysis is capable of interpreting and even predicting trends of functional changes in mutation prolific viruses.

**Author Summary**

All protein chains are compacted into globules by water pressure, as proteins have evolved as chains with alternating hydrophilic and hydrophobic segments. There are hundreds of thousands of strains of common viruses, whose full amino acid sequences are known. Here I analyze these sequences quantitatively, and show that their evolution has easily recognized Darwinian (punctuated) features. These features are global in nature, and are easily identified mathematically using global methods based on thermodynamic criticality and modern mathematics (post-Euclidean differential geometry). While the common influenza virus discussed here has been rendered almost harmless by decades of vaccination programs, the sequential decoding lessons learned here are applicable to other viruses that are emerging as powerful weapons for controlling and even curing common organ cancers. Those engineered oncolytic drugs will be discussed in future papers.

**Introduction**

In 1918 an avian-related strain (possibly associated with several years of crowded and cold trench warfare conditions) of the common influenza virus H1N1 was responsible for the largest and most lethal pandemic in modern times, with a case fatality rate of 2%. Another strain, this time supposedly swine-related, was responsible for the 2009 flu pandemic, with an 0.03% case fatality rate, nearly 100 times smaller. It was suggested [1] that antibiotics might have reduced later fatality rates associated with pneumonia, but these were not available in the 1920's, after the first pandemic disappeared in a year with a rapid drop in fatality rates. Several genes have been proposed to account for the still uncertain origins of the ability of the 1918 virus to replicate in lungs. It has been proposed that antibodies specific for conserved regions of the hemagglutinin stalk and the viral neuraminidase have caused the disappearance of more virulent H1N1 strains [2]. Much more information is available on H1N1 evolution after 1930; during this period H1N1 virulence appears to have declined, possibly as a result of Army vaccination programs beginning in 1944 [3, 4].

There are 470 amino acids in NA, and identifications of "hot spot" epitopes involving 15 or fewer amino acids responsible for the long-term decline in H1N1 virulence can be ambiguous [5,6]. Small intraspecies strain and even large interspecies dependencies [from chicken to human] of static crystalline protein structures are often too small to be resolved, even if such structures are known (which is rarely the case) [7]. Here we exploit the large number of historic H1N1 amino acid sequences available in Uniprot and NCBI databases to construct a quantitative NA sequential panorama that exhibits the effects of migrational and vaccine pressures on H1N1. The NA amino acid historical (1945-2011) panorama exhibits, perhaps for the first time, multiply punctuated evolution on a molecular scale within a single species. Apart from its specific virological applications, this panorama has important implications for the universality of the nature of protein-water interactions, especially for membrane proteins.

Before discussing details, we should examine the underlying connections between static protein structure and dynamical protein functions. One can begin by distinguishing between short- and long-range interactions. The short-range interactions often involve near or actual covalent contacts (for example, Cys-Cys disulfide bridges) or noncovalent anionic (D,Asp or E, Glu) – cationic (R, Arg or K, Lys) salt bridges mediated by H bonds. Short-range interactions are generally the only interactions discussed specifically for interactions between chemicals and proteins, and some of their energies and entropies involved can be estimated by mutagenesis and NMR experiments [8]. The protein-protein interactions important for viral membrane fusion, penetration, replication and intracellular transport involve a multitude of long-range water-mediated interactions that are separately inaccessible to direct experiment and even to accurate numerical simulations.

While the long-range interactions cannot be measured separately, one can anticipate that they will be more important than short-range interactions for viral evolution. This is because viruses are mutation prolific, and any change in short-range interactions is likely to be too large to be included among such numerous mutations. Instead these mutations will involve combinations of many small and partially compensating long-range interactions. Even in "simple" proteins like lysozyme $c$ there are many indications of compensating short-range mutations in long-range backbone flexibility between human and chicken realizations, separated by 400 million years of evolution [7,9]. If we think of viral protein structures nearly statically, long-range hydropathic interactions could dominate tertiary conformational changes between large static structural elements, such as the HA heads and/or stalk domains [2,10].

Another approach to quantifying long-range interactions involves regarding the glassy water films covering protein chains as surface chain packages constrained by water-protein interfacial tension that are deformed by the combined effects of mutational and dynamical conformational differences. If one of the latter is dominant, then it may be possible to monitor mutational trends hydropathically, through quantitative trends in solvent-accessible amino-acid specific surface areas [11,12]. There are several advantages to using hydropathic scales: (1) they are parameter-free and are independent of the protein family studied, as they are based bioinformatically on the entire Protein Data Base, and (2) the KD scale [11] describes larger changes which have a first-order or local thermodynamic character, while the MZ scale [12] has a second-order (specifically non-local) character, well adapted to describing partially compensating multiple smaller long-range interactions generated by water chain packaging forces.

Historically long-range water-protein chain interactions have been largely ignored, because the glassy surface water H bond network interactions are not easily included in molecular simulations. However, this does not mean that these small (but very numerous) interactions are unimportant; quite the contrary, they can easily connect the mutational and dynamical conformational differences, and provide much more accurate and readily available estimates [7] of property trends than backbone strain energies, which are also long-range [9].

An analogy helps one to see why this should be so. Planar aromatic hydrocarbons, such as benzene, $C_6H_6$, react with a wide variety of molecular substitutions. After the discovery of this planar family in the 1870's, chemists spent more than 60 years trying to understand their properties without much success, even after the discovery of quantum mechanics. During this period they tried to explain the preferential substitution (mutation) of peripheral H atoms by other molecules (such as methyl, $-CH_3$) at various edge positions (which are different for polycyclics). The problem was that they tried to use the in-plane or strong internal short-range σ bonds, but what they should have used was the weak surface bonds normal to the plane, called the π bonds, which are much more polarizable, with long-range interactions. In 1931 Hückel figured this out, but even then his ideas did not reach Britain and America until 1950, nearly 20 years later. Everyone was fixated on the internal σ bonds because these were obvious. The weaker π surface packaging bonds were "hidden in plain sight". This was not a blind spot of a few individuals for a short time, but a collective blind spot of all chemists for several generations. For many decades no one saw that the substitution reactions were weak and left the structure nearly planar and the internal σ bonds intact, and so the preferences surprisingly would be decided by modifying the weak π long-range surface or packaging bonds. The detailed discussion given here is more easily understood when one similarly pictures the calculations as reflecting stretching or shrinking of the water films that package protein chains from different viral strains. Water is highly polarizable (dielectric constant ~ 100 in the ms time range), and this suggests that the packaging energies can be critical.

Prolific mutations, which viruses undergo in developing new strains to evade either migrational, vaccination programmatic (influenza) or pharmaceutical (HIV) pressures, seem complex and inaccessible to engineering methods. A simpler case is the single disease mutations of rhodopsin that are responsible for *retinitis pigmentosa,* RP (retinal degeneration). Some of the methods used here to describe vaccination pressures on HA and NA (550 aa) are similar to those previously used to explain the frequency of occurrence of single RP mutations [13]. The reader who is unfamiliar with hydropathic analysis may want to study the simpler cases of lysozyme *c* (130 aa) [7] and rhodopsin (350 aa) RP [13] to see how chain hydropathic profiles using both the first-order KD and the second-order MZ scales perform in the contexts of those simpler cases. While there are many similarities, the changes in both size and functionality are also reflected in specific and instructive technical differences.

This paper also introduces some new methods specific to the large data bases associated with mutation prolific viruses. In earlier papers hierarchical hydropathic amino acid profiling was combined with sequence similarities (BLAST) to identify evolutionary optimization of blue rhodopsin (in humans) [14] and red opsin (in cats) [15]. Here BLAST amino acid similarities are maximized to construct shortest paths between apparently punctuated stages [16] of H1N1 evolution. These shortest paths can then be analyzed hydropathically to separate competing migration and vaccination effects on molecular structure at the amino acid level.

**Methods**

The second-order MZ scale $\psi_{MZ}(aa)$ is based on self-organized criticality (SOC). SOC explains power-law scaling, and it is arguably the most sophisticated concept in equilibrium and near-equilibrium thermodynamics. By identifying power-law scaling and associating it with SOC, theory explains many very widely observed non-biological phenomena [17]. The appearance of SOC in globular proteins is most spectacular, because it incorporates evolutionary optimization and makes possible many hierarchical calculations. Hierarchical calculations can be controlled and used to engineer nearly optimal properties of many systems.

The abstract concept of SOC in protein chains was proven by analyzing the properties of solvent-accessible surface areas (SASA) of 5526 protein segments from the Protein Data Bank as a function of the $C_\alpha$ segmental length $X = 2N + 1$. As N increases, internal aa overlap initially causes the SASA associated with the central segmental aa to decrease rapidly with N, but because of globular compaction, power-law behavior takes over for distant residues with $4 \leq N \leq 17$ ($9 \leq W \leq 35$)

$$dlog(SASA(aa))/dlogN = -\psi(aa) \tag{1}$$

where $\psi(aa)$ was constant for $N \geq 4$. In other words, the decrease of SASA with increasing segment size X asymptotically follows a fractal power law $N^{-\psi}$, with dimensionless exponents $\Psi = \{\psi(aa)\}$ specific to each aa. The $\Psi$ scale established by these exponents can be compared numerically to earlier non-critical hydropathic scales based on simple aa-specific areas of the entire protein (effectively $N = \infty$) or KD transfer energies (see Table I of [7], which lists all 20 amino acid constants for three $\Psi$ scales).

Readers who are unfamiliar with the concept of SOC may be surprised to find that for most proteins' functionalities the MZ $\Psi$ scale yields substantially better correlations than transference scales (such as KD), which have greater intuitive appeal. In fact, the discoverers of the MZ scale did not immediately make the connection between their results and SOC in their original paper [12], but have since adopted it as the most natural explanation for their remarkable discovery [18]. The connection of the MZ scale to SOC is valuable for several reasons. First, it provides a quantitative realization in the context of amino acid sequences alone of one of the major factors in protein evolution, global (holistic) water-protein chain packaging interactions, because evolution tends to optimize these interactions for homologous chain folds. While evolution has refined these interactions, they are still not perfectly ideal (the critical point for a given fold is approached but not attained). Second, water-protein chain packaging interactions are of course weaker than direct aa steric interactions at short range, which is why the self-similarity power law (1) holds only for $N \geq 4$. (Notice that the natural pitch of the α helix is 3.6.) Third, it is clear that transference scales like KD are thermodynamically similar to first-order phase transitions, whereas SOC is similar to a second-order phase transition. This distinction is a valuable one, because it means that which scale gives better correlations will probably depend on whether the function analyzed involves a few large (first-order, partially unfolding) interactions, or many small (second-order, small relative tertiary, fold-retaining) interactions, or possibly the two combined in series. In all the homologous cases we have studied so far, including not only rhodopsin, but also a number of other GPCR signaling proteins [19] as well as lysozyme *c* and many nucleoporin repeat transport proteins [20], the thermodynamically second-order nonlocal MZ Ψ distal scale based on SOC has outperformed first-order local transference hydropathic scales.

One can define the nonlocal sliding window average by

$$<\psi(j)W> = \Sigma \ \psi(j + i)/W \quad (2)$$

with $-N \leq i \leq N$. Sliding window averages are not new to proteins; they are mentioned in the titles/abstracts of 100 + papers. It turned out that significant differences in the hydropathic sliding window average $<\psi W>$ (W = window length) of lysozyme *c* (a non-membrane protein) between species occurred only in certain long (W ~ 15) segments in the SOC range, which could be identified by using $<\psi 3>$ MZ chain profiles. These segments (for instance, the newly identified enzymatic blue glycosidic scissors, aa 80-94, apparent but not identified functionally in lysozyme *c* backbone strain studies [9]) correlate well with evolutionary trends in both enzymatic and antimicrobial properties [7]. These correlations probably arise from dynamical tertiary conformational mechanisms. They are not explicable using most other tools, such as the $<\psi 3>$ KD chain profiles, BLAST sequential similarity or conventional backbone Euclidean geometrical structural superpositions. Some similarities can be seen between [7] (hydropathic chain profile evolution) and [9] (mutational changes in backbone flexibility), as both identify aa 80-94 as dynamically important. However, [7] established this segment as enzymatically crucial, while [9] merely recognized it as representing a narrow mesa of increased mutational flexibility, without mentioning its enzymatic functionality.

Here we will see that hydropathic elastic chain roughening profiles provide a tunable nonlocal handle on long-range MZ interactions, which could be as useful for the latter as short sequence motifs and crystal structures are for exploring short-range (contact) interactions. Roughening is defined simply in terms of the variances of hydropathic $<\psi W>$ sliding window chain profiles, where the window length W can be tuned to optimize resolution of interprotein (for instance, different species or mutational) differences. Generalized hydropathic chain convoluted profiles $<\psi_\alpha(i)W>$ exhibit oscillations, and these oscillations are often quite similar for different species over large parts of a given protein. In retrospect these strong similarities could have been expected, as the conformational changes that determine protein functionality involve tertiary long-range forces, with the changes in short-range interactions limited by rigid secondary structures (helices and strands) whose main role is to stabilize the functional units during long-range conformational changes. Increasing the convolution length W smoothes the oscillations of the hydrophobicity chain profiles and reduces their amplitudes, and yields parameter-free measures of long-range interfacial water-protein roughening. The variance function $R(W)$ is defined by

$$R(W) = \sigma^2(W) = \Sigma_i \ (<\psi(i)W(N)> - <<\psi(i)W(N)>>)^2/(M-W) \qquad (3)$$

where the sequence contains M aa, and the average is over the central sites, with N sites at each end excluded. Note that because of Eqn. (1), $\langle\psi(i)W(N)\rangle$ is in effect a double W-X convolution, as $\psi(i)$ itself is based on the linear region of a log-log plot of SASA as a function of X = 2N+1. The variance function R(W) measures stretching or shrinking of the water films that package different viral strains.

**Paths in Hydropathic Configuration Space**

As we will see, hydropathic scales enable us to compare homologous amino acid sequences. For a protein with 500 amino acids, each of which is selected from a list of 20 possibilities, this means that the maximal dimension of the corresponding configuration space is 10,000. To reduce evolution in this space to a manageable form, one can construct shortest paths from sequence A to sequence B that involve minimum numbers of mutations of intermediate strains. This procedure is especially simple if the strains are ordered chronologically, but they can also be ordered according to a property such as roughness. Mathematicians will recognize such paths as analogous to the method of steepest descent, originated by Laplace, extended to saddle points by Riemann (1863) and Debye (1909), and discussed in many Web-accessible lectures.

**Tuning W**

What is the meaning of tuning W? Increasing W from 1 to some larger value of 2M + 1 (M > 0) does have a smoothing effect, but how does W relate to structural features? The most prominent feature of NA chain hydroprofiles is the hydrophobic peak associated with the 7-35 transmembrane region, so W should be small enough to resolve this peak (other peaks and valleys have similar widths, as do the glycosylation sites and/or disulfide bond spacing, as shown on Uniprot). At the same time, the difference between $R(W)$ calculated by the first-order KD scale and the second-order SOC MZ scale increases with increasing W, and we would like this difference to be as large as possible. For these two reasons, we chose $W = W_{max} = 17$, and this choice has worked well. Similar results would probably be obtained with W between 15 and 21. One can use the hydropathic tools (1) – (3) inductively, following the author's own exploratory path based on extracting as much as is easily possible bioinformatically, using primarily amino acid sequences of NA as they have evolved from 1918 to 2010. The following NA results are presented using two scales, the first-order like KD scale, and the second-order like MZ SOC scale. Generally the roughnesses calculated from the KD scale are larger than those from the MZ scale for both HA and NA. It seems likely that this reflects the greater accuracy of the MZ scale, because viruses evolve towards smoother chain profiles, and some features of this evolution are unresolved with the KD scale.

**Results**

Widespread vaccination programs can interrupt the stability of similar viral strains, which can be called punctuated equilibrium [16]. Here we prefer the term punctuated evolution, which has a different meaning [21]. Hitherto efforts to study punctuated molecular evolution have focused mainly on microorganisms [22], but our results exhibit punctuated evolution for $R_{KD}(W_{max})$ and $R_{MZ}(W_{max})$ in human NA H1N1, as initiated by migration (interspecies or geographical intraspecies) and vaccination pressures.

Punctuated evolution is a more quantitative concept than directed evolution [23]; it is reminiscent of diffusive foraging in complex network systems, which often can be described as a product of many small (equilibrium) and a few large (evolutionary) Lévy hops [24]. Our results on the evolution of NA H1N1 are summarized in Table I, which exhibits sharp punctuations, both favorable (vaccination) and unfavorable (migration).

Flu vaccine was first widely administered by the American Army starting in 1944. The pressure created by vaccination of < 0.1% of the world's population drastically smoothed NA H1N1 chain profiles $R_{KD}(W_{max})$ and $R_{MZ}(W_{max})$ almost everywhere (Fig. 1). Overall the KD and MZ values appear to show similar trends, but there may be less scatter in the MZ values. The broad trend is nearly constant from 1918 to 1948, superstrain A [(MZ,KD) ~ (164,211)], when the cumulative effects of the Army vaccination program led to a punctuated drop of $R_{KD}(W_{max})$ by about 10% and of $R_{MZ}(W_{max})$ by about 13%, first unambiguously observed in 1950 Fort Warren, followed by the roughness "vaccination valley", which lasted until 1976, superstrain B [(MZ,KD) ~ (148,188)]. Note that because of its geopolitical isolation the 1954 Leningrad strain [(MZ,KD) = (165,204)] was apparently unaffected by the Western vaccination program. A curious feature of this period is the two strains found in the Netherlands in 1948 [(MZ,KD) ~ (183,222)], and five years later, 1953 [(MZ,KD) ~ (145,188)], with a punctuated gain in smoothness twice as large as the average for other strains.

Let us take a closer look (Fig. 2) at the onset of the "vaccination valley" between 1945 and 1950. It occurred abruptly ("punctuated"), with a large drop between 1949 Rome superstrain A [(MZ,KD) ~ (164,211)], and 1950 Fort Warren superstrain B [(MZ,KD) ~ (148,192)]. In the 1945-1949 period, although $R_{MZ}(W_{max})$ appears to be nearly level, from one year to the next out of 470 amino acids there were many fluctuations in superstrain A (using the BLAST classification, typically ~ 35 mutations, including 15 non-positive, that is, belonging to separate groups (hydrophobic, phylic or neutral). However, after 1949 these fluctuations almost disappeared, with only ~ 15 mutations, < 5 non-positive in superstrain B. As the strains became smoother, they also became more stable. Although this effect is easily recognized afterwards with BLAST, it has been found here only because we were aware of the $R_{MZ}(W_{max})$ punctuation between 1949 Rome and 1950 Fort Warren.

A still closer look at the 1949 Rome and 1950 Fort Warren punctuation is shown in Fig. 3. Between Houston 1945 and Rome 1949 there were 11 largest mutations in superstrain A, defined as involving changes in MZ hydropathicity $\psi$ of > 50 (in units of $10^{-3}$). Substitutions near the N terminal, especially V16A, reduced $R_{MZ}(W_{max})$, but then other mutations (esp. I362T) near the P terminal compensated these, so that $R_{MZ}(W_{max})$ was only slightly reduced on going from Houston 1945 to Rome 1949. The actual punctuation occurred between 1949 Rome and 1950 Fort Warren, and it involved only 5 largest mutations, with only one of these (I19T) having a large uncompensated effect, opposite in sign to the prior I362T, and nearly all the largest mutations reducing $R_{MZ}(W_{max})$. This punctuated change was caused mainly by the single mutation I19T. If applied alone to Rome 1949, it causes $R_{MZ}(17)$ to drop from 164 to 153, about 2/3 of the $R_{MZ}(17)$ drop to 1950 Fort Warren ($R_{MZ}(17)$) = 148).

These apparently small "vaccination valley" A-B punctuated shifts in NA smoothness were accompanied by large reductions in antigenic response, so much so that interest in H1N1 rapidly declined during the early 1950's, and a gap appears in the H1N1 sequence data base, extending from 1958 to 1975.  (Meanwhile, other unusual subtypes appeared, such as H2N2 in 1957, and H3N2 in 1968, which attracted much attention.)  Attention was refocused on common H1N1 when it reappeared in a strongly virulent AM form in Jan. 1976 in Fort Dix in a mysterious and surprisingly brief outbreak that lasted only 3 weeks and did not spread beyond Fort Dix. Experts were unable to answer many questions, "including the following: Where did A/New Jersey (Fort Dix) come from? Why did transmission stop?" [25].

With modern data bases these historic questions can be answered.  ACU80017 (A/Fort Dix 1976) is a throwback strain identical to ADJ40425 (Puerto Rico 1934).   The "new" Fort Dix strain came from basic trainees imported from Puerto Rico, a subtropical island sufficiently isolated from temperate climates that its rural inhabitants could have evaded the beneficial effects of vaccination, with the same strain persisting for more than 40 years.  Transmission stopped because of prompt corrections to crowded and cold basic trainee conditions.  Although the Fort Dix strain was labeled possible swine flu, it could also have been simply a pre-vaccination form of H1N1, possibly persisting only in rural areas, and subtropically more antigenic than earlier temperate strains.

To the officials who handled the Fort Dix outbreak it appeared that the incipient epidemic had been contained, but as Fig. 2 shows, this was not entirely so.  In response to the Fort Dix outbreak, 25% of the American population was vaccinated by October 1976, using a hastily produced vaccine that does not meet modern standards of purity [26], which may have had a few unfortunate side effects.  However, as shown in Fig. 2, in spite of massive vaccinations, after 1976 the earlier roughness of superstrain A (before the vaccination valley of 1950 – 1975, and matching the level found in Russia, also isolated from vaccination pressures) reappeared as strain A´.  Between 1950 and 1975, the antigenic level of flu was lower [the superstrain B 'vaccination valley"]; the retrograde return to prior superstrain A levels in 1977 in strain A´ is especially clear using the MZ scale.

   The expected  "Fort Dix" epidemic did not occur, and it is widely believed that the hasty vaccination program was merely a big "fire drill"  [26].   The data shown in Fig. 2 suggest a more complex pattern.  By the 1970's more than 1 million Puerto Ricans had emigrated to the mainland United States, where they were concentrated in neighborhoods of larger cities, and frequently visited relatives in Puerto Rico.  It appears from Fig. 2 that the Fort Dix strain was a more antigenic strain than strains before 1950.  The strains that circulated in the late 1970's and 1980's were antigenically similar to the strains before 1950, possibly because the Puerto Rican migration pressure was balanced by the 1976 vaccination pressure.  Had the 1976 massive vaccinations not occurred, the highly antigenic Fort Dix H1N1 level might well have occurred in a worldwide pandemic.   Similar but weaker conclusions were reached in a recent mouse challenge study [27], an experimental analogue to the present superstrain hydroanalysis.

   NA punctuated superstrain B-C evolution also occurred mainly in 1987[(MZ,KD) ~ (165,214)]-1989 [(MZ,KD) ~ (149,201)], but it went unnoticed, probably because it was the result of the gradual accumulation of vaccination benefits.   However, with the MZ scale one can study this B-C transition and find strains that represent the crossover explicitly.  One can also find the single mutations that caused the largest smoothing.  Although there are 15 non-positive mutations between Singapore 1986 [(MZ,KD) ~ (166,212)] and Bayern 1995 [(MZ,KD) ~ (141,200)], about half of the smoothing is due to a single key transmembrane mutation, C14S.  This mutation first appeared in Japan, for instance in Hokkaido 1988 and Yamagata 1989 [(MZ,KD) ~ (151,210)].  Apparently by the mid-1980's the vaccination program in Japan among schoolchildren begun immediately after Fort Dix in 1976 [28] had become exceptionally effective, and its benefits gradually spread to the rest of the world. The smoothing increase is clearer with the MZ scale than with the KD scale, because of the large scatter in the latter's results, evident in Table I. Note that the single mutation most responsible for 1949-1950 punctuated smoothing, I19T, also lies near the center of the transmembrane region 7-35 (Uniprot); both replace a TM central hydrophobic amino acid with a hydrophilic one, making the TM region more flexible.

Most of the post-1988 mutations also refined the smoothness of the 1977-1986 strain until a milder superstrain C, better able to evade vaccine, appeared and by 1996 the full effects of vaccine avoidance gave superstrain C [(MZ,KD) ~ (142,199)]  Because superstrain C was much milder, fewer vaccinations occurred, and simultaneously there was a steady increase in migration.  The first appearance of "swine flu" was in a Beijing 1995 strain (ACF41870) [(MZ,KD) = (154,205)], with an MZ roughness $6\sigma$ above the superstrain C average, and presumably severe symptoms, similar to those not seen since before 1988.  This observation led to measurements of 23 NA amino acid strain sequences in New York in 1995.  The earliest of these in 1995 were near the 1988-2005 superstrain C averages, but the latest in 1995 showed an increase of MZ roughness of $\sigma$, and by 1996 this had increased further to $1.5\sigma$.  These MZ roughness increases were significant, but on the scale of the difference ~ $10\sigma$ between 1977-1986 and 1988-2005 superstrain B and C averages, these effects did not constitute a significant threat to public health.  Moreover, multiple 1996 measurements in Memphis (9) and Nanchang (13) showed no or little increase in MZ roughness and presumably symptom severity. It appeared that the Beijing 1995 strain was an aberration, which had had small effects in New York, contained by vaccination counter-pressures.  However, another 1977-1986-like strain appeared in Hong Kong in 2002 (ACA96509) [(MZ,KD) = (158,213)], indicating that more severe "throwback" strains could re-emerge and flourish in crowded conditions.

Signs of more severe flu were widespread in New York by 2003 and in German cities by 2005, leading to measurements of 15 NA amino acid strain sequences.  Berlin reported the largest increase in MZ roughness, about $2\sigma$ above the 1988-2005 superstrain C average, possibly reflecting more crowded conditions among migrant labor.  (Note that a direct connection between these trends and actual swine flu is not easily established by phylogenetic analysis [29,30], which appears to lack the resolution needed: it could not separate American from Eurasian swine influences.)  By 2006, the Berlin value [(MZ,KD) ~ (147,195)] had risen further to $3\sigma$ above the 1988-2005 superstrain C MZ average, and increases of $\sigma$ or more are common in the ~50 American strains reported in 2006, with Los Angeles showing $2\sigma$.  It is striking that these urban trends appear consistently using the MZ scale, with its small $\sigma$, but are not apparent with the KD scale, where $\sigma$ is more than twice as large.

Between 2005 and 2008, the number of NA H1N1 annually reported sequences in the NCBI database grew from 20 to 80, indicative of growing concern for the possibility of a "swine flu" pandemic.  In 2007 most large urban centers show larger MZ roughness, about $2\text{-}3\sigma$ above the 1988-2005 superstrain C average, with Houston (ACA33539, 147.3) reporting the largest increase, bringing it close to superstrain B.  However, New York (ABW23371,136.6), where vaccination had become common after the 2003-2005 seasons, showed beneficial vaccination effects, with 2007 MZ roughness $2\sigma$ below the 1988-2005 superstrain C average.

The most interesting 2007 cases occurred in Hawaii: they offer the opportunity to test the punctuated superstrain concepts and the effects of vaccination in apparently the same place and time.  In 2007 Hawaii reported 23 sequences, of which 10 were independent.  Using the MZ scale, these 10 sequences separate clearly into three superstrain groups (MZ;KD) ~ (148,136,130:192,194,183).  The sum $\Sigma$ of the standard deviations $\sigma$ for the first two MZ groups I {5 members} and II {4} is only 3 on both scales. Thus the separation of the I and II MZ groups is a comfortable $\alpha\Sigma$, with $\alpha = 3.8$, but these I and II groups are separated only by $\alpha < 1$ on the KD scale.  In fact, using the MZ scale only, we can identify I as tourists (for instance, from Brisbane, California, or Texas) who had not benefitted from the widespread vaccination program of New York and whose NA had regressed to superstrain B, whereas II was New Yorkers who had benefitted, perhaps even as part of their travel plan.  Note that this is a public health benefit: both I and II tourists presumably enjoyed similarly comfortable, uncrowded conditions at home. The third ultrasmooth Hawaii 2007group III [(MZ,KD) = (130,183)] had only one example, which appears to be closer to strain D below.

An interesting footnote here is that the splitting of the Hawaii 2007 I and II tourist groups in units of $\Sigma$ is a sensitive test of the choice of our tuning parameter $W_{max}$, which was set at 17 and gave $\alpha = 3.8$.  Repeating the Hawaii I and II group analysis with $W_{max} = 15$, one finds $\alpha = 2.5$ (poorer resolution), but for $W_{max} = 21$, one finds $\alpha = 4.0$ (slightly better resolution).  The improvement in $\alpha$ from $W_{max} = 17$ to $W_{max} = 21$ is so small as not to warrant repeating the $W_{max} = 17$ calculations for $W_{max} = 21$.  The poorer resolution for $W_{max} = 15$ already reflects the masking effects of short-range forces (packing misfit), which obscure the favorable effects of vaccination pressure in smoothing the large-scale, water-packaged tertiary NA structure.  These large-scale water-protein interactions are invisible to conventional similarity analysis by BLAST. It is also interesting that in studying the evolution of enzyme activity in lysozyme $c$ ( a smaller protein, only 130 aa), we identified a W = 15 aa segment (the 'blue scissors", 80-94) significantly different between humans and swine [7,9], correlated with changes in enzyme activity.

A new H1N1 vaccine for 2008-2009 was adopted by WHO, based on Brisbane 2007 (ADE28752) [MZ = 146.5, 2.5$\sigma$ above superstrain C average], and widely distributed, with vaccine avoidance beneficial smoothing effects leading to superstrain D [MZ = 125] far, far larger than had been anticipated. Not only did superstrain D reverse the urban trend toward larger roughness in the absence of vaccination, but also the wide vaccine distribution even caused the average roughness to drop about 8$\sigma$ [1988-2005 superstrain C units] below the previous level (Table I) to (MZ,KD) ~ (125,175). Given that increases of order (1-2)$\sigma$ were obvious enough clinically to launch nearly two hundred sequencing studies in 2005-2008, after there were only two in 2004, it is clear that this reduction represents a very large gain in mildness of common flu.

By the time we reach 2009, concern over the "swine flu" pandemic was such that NCBI lists more than 6000 "neuraminidase H1N1 2009 human" sequences. However, because of vaccination pressures, all these 2009 strains not only were the mildest (smoothest) known, but also their fluctuations were very small. A survey of 10 geographically dispersed (globalized) strains gave the convergent results shown compactly in Table I, where $\sigma$(MZ) is five times smaller in 2009 than in the 2005 plateau. The small values of $\sigma$ quoted here still do not reflect how small the fluctuations have become. As remarked previously, while evolution has refined protein interactions, they are still not perfectly ideal (the critical point for a given fold is approached but not attained). However, under severe vaccination pressure, virus superstrain D has become so close to perfect that the same 2009 strain [$R_{MZ}(17) = 125.5$] can be found in cities as remote as Shanghai, Bayern, Mexico City, and Singapore.  The remaining scatter comes from smaller and more isolated countries like Santo Domingo, but even it is small (two mutations, but no non-positives).  It is difficult to see how this remarkable "forced" convergence can be explained by a theory not involving SOC.  This kind of convergence is typical of SOC [31].  Documented cases of convergent biomolecular evolution due to selection are fairly unusual, and examples to date have involved only a few amino acid positions in primitive species [32].  Many more examples of convergence may be obtainable by hydropathic sequential analysis.  The interface of protein structural biology, protein biophysics, molecular evolution, and molecular population genetics forms the foundations for a mechanistic understanding of many aspects of protein biochemistry [33].

Suppose we had evaluated segmental smoothness, instead of whole-protein smoothness?  We did a few calculations of this kind, for instance on NA sites 9-81 and 9-110.  Segments are expected to be much rougher than the whole protein, and so it is.  Moreover, the roughness trends for 9-81 (the "stalk" region alone) do not parallel those of the whole protein, but at least qualitatively those for 9-110 (the "stalk" region + a stabilizing glycosylation connecting region) are similar.  (However, for reassorted chimeric viruses involving heads and stalks from two different viruses [34], $R_{KD}(W_{max})$ can increase while $R_{MZ}(W_{max})$ decreases, even for whole proteins, which is ambiguous.)  In any case the striking SOC trends exhibited by whole viruses under vaccination pressures are not expected to be nearly so pronounced for segments as for whole proteins, and neither segments nor chimerics were pursued further.

Another striking feature of the superstrain C-D (2007-2009) punctuation is that from 2005 Berlin ACI32764 or 2007 ADE28752 Brisbane to 2009 consensus, ACV42020, the number of BLAST non-positive mutations forced by very widespread vaccination is very large ~ 50, more than three times the number involved in the superstrain A-B (1949-1950) and B-C (1986-1988) punctuations. Confining attention to changes in $\psi_{MZ}$(aa) > 50, the C-D list is reduced to 20; it contains several surprises. If we think of hydrophobic to hydrophilic substations as charge transfer, then XN1Z and its inverse ZN2X are like an N1-N2 hydropathic dipole, while two such dipoles are like a hydropathic quadrupole. There are two such distinct C-D dipoles and one quadrupole. Putting them aside for the moment, we are left with 12 large changes in $\psi_{MZ}$(aa). While for the A-B and B-C punctuations the largest smoothing mutations replaced hydrophobic aa by hydrophilic aa (making the protein globule more open), here the opposite is true: there are only 5 phobic-philic substitutions, and 7 philic-phobic ones (making the proteins globule more closed). The latter even includes S14C, which reverses the key mutation of the B-C punctuation.

So how did superstrain D become smoother than superstrain C? To answer this question we can look at Fig. 4 (the C-D punctuation chain profiles) and compare it with Fig. 1 (the A-B punctuation chain profiles). The qualitative differences are large and obvious, even though the A-B and C-D decreases in roughness $R_{MZ}$(17) are similar in magnitude. In both the A-B and B-C punctuations the driving forces are only a few reductions (increases) in hydrophobic peak (hydrophilic valley) extremes, caused especially by single mutations in the 7-35 transmembrane range, softening the TM segment. In the C-D punctuation the reverse unexpectedly happens: hydrophobic peaks are enhanced and hydrophilic canyons reduced in the terminal ranges 9-230 and 340-461, and are compensated mainly by softening and smoothing in the mid-range 240-300. Intuitively one would not think the mid-range smoothing would be able to compensate the roughening of the termini, but one's intuition is probably insufficient for calculating differential interfacial water-protein chain packaging energies, which is why quantitative calculations of $R_{MZ}(W_{max})$ are so useful. Note here that softening of the chain midrange is also mechanically advantageous, as it facilitates bending of the terminal segments relative to each other in scissors-like tertiary chain domain motions φ that could be relevant to quaternary viral chain clustering. The hydrodomain bending coordinate φ can be regarded as a secondary configuration coordinate supplementing hydroroughness $R_{MZ}(W_{max})$, the primary configuration coordinate, but still subordinate to it.

The effect of the paired transmembrane C-D dipole mutation I20A and A34I on Brisbane 2007 $R_{MZ}(17) = 146.5$ is large (reduction to 141.7), while the closely spaced pair T46I and I48T reduces $R_{MZ}(17)$ only to 145.9 (8x smaller reduction). Note that the separation of the latter is only two sites, while the former is separated by 14 sites, corresponding to a 7x larger dipole moment "arm". The quadrupole mutation V75A, A166V, V232A and A454V reduction is also large (141.8). All three shifts are reductions, taking C towards D, with no change in overall hydrophobicity.

Were the mid-2000 strains imported from swine? It appears that they were, although an earlier attempt to answer this question genomically failed [35] because of genetic drift. A BLAST search on swine (Minn.) 2007 yielded many human strains 2003-2007 (superstrain C) in a similarity range less than twice as wide as the swine range itself (well-mixed swine and human strains). The swine 2007 profile, plotted in Fig. 4 together with the human 2003-2007 strains, is nearly indistinguishable from Berlin 2005, New York 2003, etc. However, a similar search on swine (Italy) 2000 did not yield any human strains over a range more than twice as wide as the 2007 swine range (no contemporary mixing), and then the nearest human strains all dated from 1976 or earlier (superstrain B). It appears that the swine and human strains had converged (beginning around 2003 in New York, 2005 in Berlin, etc.) on superstrain C, mainly because of swine flu drift. The effects of the swine flu vaccination program shown in Fig. 4 on human NA are thus quite remarkable, as the human strains made a very large "Lévy" jump [24] from superstrain C to superstrain D to avoid the vaccine and dodge swine flu, while reducing their severity. Without the swine flu vaccination program it seems most likely that both human and swine flu would have continued to drift back to the strain B levels or even the strain A severity levels. Thus the vaccination program had a striking and unprecedented adaptive effect.

The 2009 NA sequences are dominated by a single strain, which appeared around the world in several thousand identical reported copies. The second most popular strain occurred nearly ten times less frequently, and it involved only two positive mutations. From 2003 to 2007, before the swine flu vaccination program, the single most popular strain occurred about half as frequently as in 2009. This shows most effectively the convergent influence of vaccination pressures. There were fewer reported copies of the most popular 2009 strain in 2010, showing a modest relaxation of vaccination pressures. However, there is no evidence so far in the 2010 and 2011 sequences for significant increases in roughness due to migration divergence or relaxation of vaccination pressure. One interesting point is that vaccination pressure has apparently been sustained in Washington DC, with the result that strains reported there in 2011 were dominated by a single strain D*, which reached the lowest roughness reported yet, $R_{MZ}(17) = 120.7$. Once again this D-D* reduction is dominated by a single mutation, S339L. which increases hydrophobicity and further raises the deep valley near 340 in Fig. 4, thereby reducing $R_{MZ}(17)$.

The existence of D* raises an interesting question:  could we engineer a strain to be even smoother than D* and still be replication competent?  An obvious way to do this is to bring back C14S.  In so doing we might lose some of the secondary scissors-like tertiary activity φ that characterizes D compared to C, but the possibility is certainly worth testing.  To do it consistently (without disrupting short-range interactions and introducing primary or secondary packing misfit), we should splice C with D* in a primer range of ~ 30aa where two partial sequences are conserved.  Comparing C (Bayern1995) with D* (Washington DC 2011) we see such a range near site 200, so the splice is 1-200 Bayern1995 – 201-470 Washington DC 2011, which gives $R_{MZ}(17)$ of 117.6.  This is a modest improvement on D*, but still significant compared to D, where $R_{MZ}(17) \sim 125$. There is also a 25 aa primer range near 100 that could be used, which gives $R_{MZ}(17) = 114.4$.  Thus a splice could provide beneficial effects for moderating influenza illness through introduction of a new superstrain.  It could be described as superstrain E.

**Discussion**

Viral evolution is conventionally discussed in terms of antigenic drift, as displayed by antigenic cartography (AC) [36].  AC relies mainly on measuring strain separations in terms of numbers of amino acid substitutions, and converts these numbers into two-dimensional maps by iterative triangulation (pre-Euclidean land surveying, > 5000 years old); AC represents an improvement on previous efforts [35].  These aa similarity maps are then compared to triangulated maps based on measuring inhibition titers to quantify free sialic acid, and a few strains are examined for qualitative similarities of sialic acid maps to amino acid similarity maps. This approach, which relies entirely on map similarities of similarity maps, is undirected and does not appear to be capable of recognizing punctuated hierarchical migration/vaccination evolution quantitatively.   It did identify a "noncontinuous pattern of NA drift in the last 15 years" [1995-2010], which it assigned to short-range ionic interaction reversal primarily associated with E329K. According to Fig. 4, the hydrophilic minimum in the NA hydrochain  occurs near 330, so E329K could be important, although it has only a small hydroeffect.  This could be some kind of punctuation, so we checked our 10 Hawaii examples and found an excellent correlation: Group I {5} had only 329E, II{4} had three 329K and one "false" 329E (ACA33652), and III had 329K (continuation of II).

Previously we suggested that group I could have come from Brisbane, California, New York or Texas, but in 2007 only Brisbane had 329E, so Brisbane was probably the source of the NA reversion towards superstrain B roughness seen in Hawaii in 2007 group I. Most of the other strains examined by us had 329K, including strains as far back as Melbourne 1935.  A few 2011 strains still have 329E or D, while the most evolved and smoothest Wash DC 2010 strains have 329N (neutralized 329D).  The effect of K329E on the difference between 2007 Brisbane and Bayern 2009 in Fig. 4 is limited to only one of many hydrophilic extrema, and it is not an important part of the superstrain C-D  roughness punctuation shown in Table I and Fig. 4.  The available smoothness trends suggest that 329E was but one of the ~ 17 mutations among pairs of 2007 NA strains.  Significant mutations also occur nearby at 331 and 332, with K329N occurring together with G331K (charge compensation pair mutation).  Altogether 329E Brisbane 2007 appears to be only a passing aberration of importance secondary to migration-driven globular roughening, which in turn has led to overriding vaccination-driven convergent globular smoothing.

More sophisticated sequenced-based methods of measuring antigenic distances for influenza viruses have been explored by Deem et al. in a series of papers involving, for example, identification of clusters of 2009 H1N1 strains, connected according to time-ordered single amino acid mutations and minimal nucleotide mutations [37].   This approach has proved to be at least as predictive of human vaccine effectiveness as HA inhibition assay data from ferret studies.  Their sequence analysis has been extended to include antibody surface epitopes and charge interactions [38] in H3N2 free energies.  It is possible that charge interactions are exceptionally dominant in H3N2, but not in H1N1.  (This would be consistent with the dominance of H1N1 among circulating flu viruses.)  However, as was discussed in the Introduction, we believe that including evolution through SOC water film packaging as embodied in the MZ hydropathicity scale is normally a more accurate and neutral way of approaching surface interactions, and this appears to have been confirmed by our identification of punctuated convergence of NA1.  Neither mutational nor distance measurements appear to identify kinetically critical mutations quantitatively, or to separate N-glycan and sialic acid interactions.

Here we have used hydroanalysis to establish an historical (1945-2011) panorama of vaccination- and migration-driven punctuated mutations.  In NA most of these punctuations have been driven by mutations of a few key amino acids, which our analysis has identified.  The most recent NA punctuation, a vaccination-driven reaction against swine flu, is different, and involves large collective effects and many mutations, which smooth the central receptor region associated with sialic acid binding [39].  Simple overall hydropathic smoothing by a few mutations (early punctuations) can accelerate viral membrane fusion, while central region smoothing (C-D-D*) could facilitate conformational changes which could cause NA to fold back upon itself [40].

Our central conclusion is that large-scale vaccination programs have been responsible for a very large reduction in influenza severity in the last century, initially by inducing a few key mutations to superstrains A and then B and C, but most recently by driving a large-scale jump from C to new and remarkably compact, mild and convergent superstrains D and D*. These programs have been stimulated by public concern over increases in severity due to mixing with other species, migration, or by a desire to protect children or the aged [28]. The long-term value of vaccination programs is not easily quantified by personal experience or even by large scale but only short-term analysis [41]. However, the succession of long-term vaccination plateaus found here leaves little doubt that even relatively small programs (like Japanese children, beginning 1976, involving ~ 0.1 -1% of the world's population) have had large and lasting beneficial effects for all, much more so than indicated by short-term studies [41]. In this respect the ongoing world-wide success of limited vaccination programs is most impressive.

One can only speculate concerning the mechanism underlying such large collective benefits from so relatively small vaccination programs. One possible explanation is that there was a large, geographically dispersed pool of mutated viruses, and that vaccination pressures against more severe strains exploited the superior replication and transmissibility of smoother and less severe superstrains, at the same time reducing the size of the pool. Globalization (for instance, intercontinental jet travel [24]) and urbanization also appear to have contributed to the convergence of mutated strains by 2009, with vaccination programs producing "reverse pandemics".

The central limitation of prior studies of viral kinetics has been their low resolution, limited to the N-glycan length scale. There one finds evidence that N-glycans guide partner ligands to their binding sites and prevent irregular protein aggregation by covering oligomerization sites away from the ligand-binding site [42]. Here we have shown that detailed hydropathic chain profile smoothness analysis of NA enables resolution at the level of individual glycoprotein amino acids. Why is increasing globular "ball bearing" smoothness, which facilitates granular rotation during shear flow [43], so important kinetically at the molecular level? Formation of oncolytic core oligomers requires multiple steps. First, individual viruses must be bound to the cancer cell membrane. Next these isolated molecules must diffuse along the membrane surface to form clusters. The surface diffusion rate over a rough surface can be accelerated by smoothing both the NA and HA glycoproteins, another reason for viral glycosidic balance [44].

The success of smoothness analysis and our historical (1945-2011) panorama depends almost entirely on the large viral genetic data base. The data have come from many sources, but the largest part of this data base, especially the older parts, is due to the NIAID Influenza Genome Sequencing Project, which has proved to be invaluable. The present analysis has important implications for engineering oncolytically superior strains of F and HN glycoproteins of Newcastle disease virus for cancer therapy [45], which will be presented separately, as will an analysis of HA.

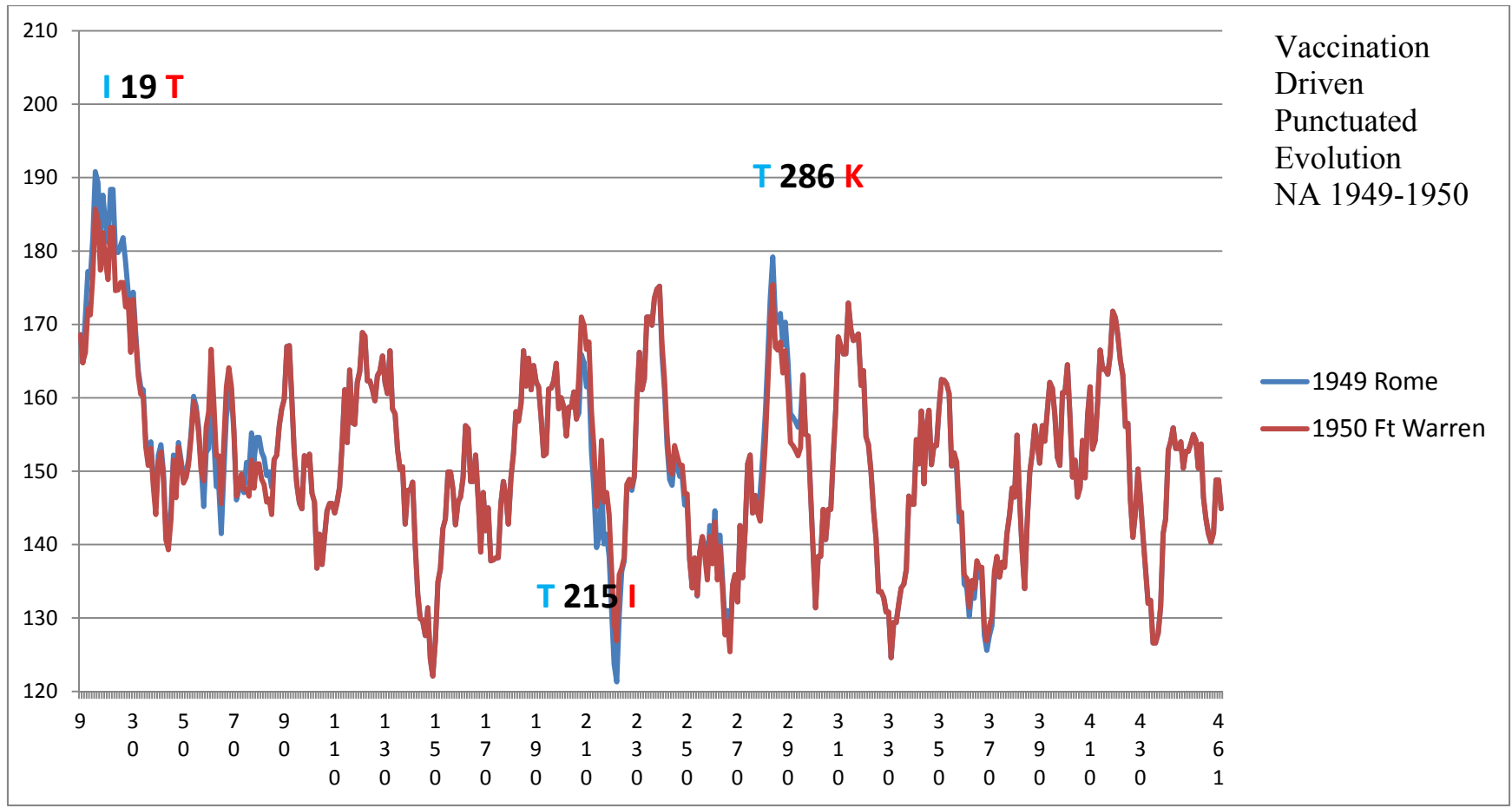

Fig. 1. The NA hydrochain s <ψ(j)W> with W = 17 are shown before and after the 1950 vaccination punctuation (A-B in Table I). The largest NA changes due to three single mutations are indicated. These changes look very small here, but their effects on roughness are much larger than the scatter σ in the roughness averages over many strains in each superstrain plateau. This shows how

informative the $R_{KD}(W_{max})$ and $R_{MZ}(W_{max})$ are (especially $R_{MZ}(W_{max})$, because of its smaller σ's), and why they are so successful in measuring vaccination and migration pressures.  Note how I19T is centrally placed in the overall strongly hydrophobic transmembrane region 7-35. Overall the changes are concentrated near extrema, and smooth the chain profile globally by lowering peaks (like 19) and raising valleys (like 215).

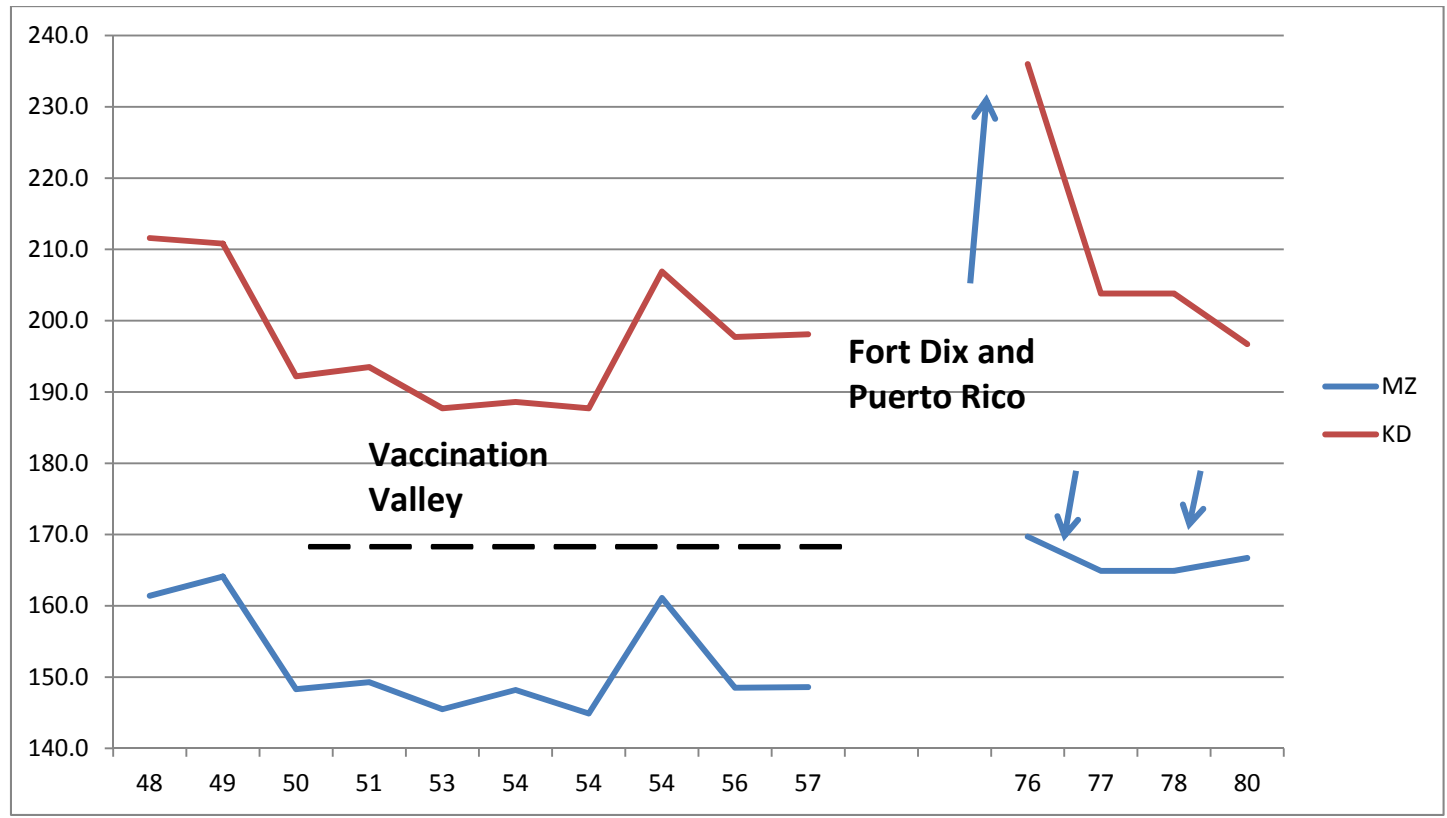

Fig. 2.  A sketch showing the NA17 evolution of both MZ and KD roughnesses through the first vaccination and migration punctuations.  The horizontal dashed line draws attention to the reversion after the  6-week Fort Dix outbreak of world-wide strains to the superstrain A plateau prevalent before the success of the Army vaccination program, which had led to the vaccination valley.  The 1954 value refers to an isolated Russian strain (see text). Otherwise, example species were sometimes selected by the steepest descent methods discussed in text.

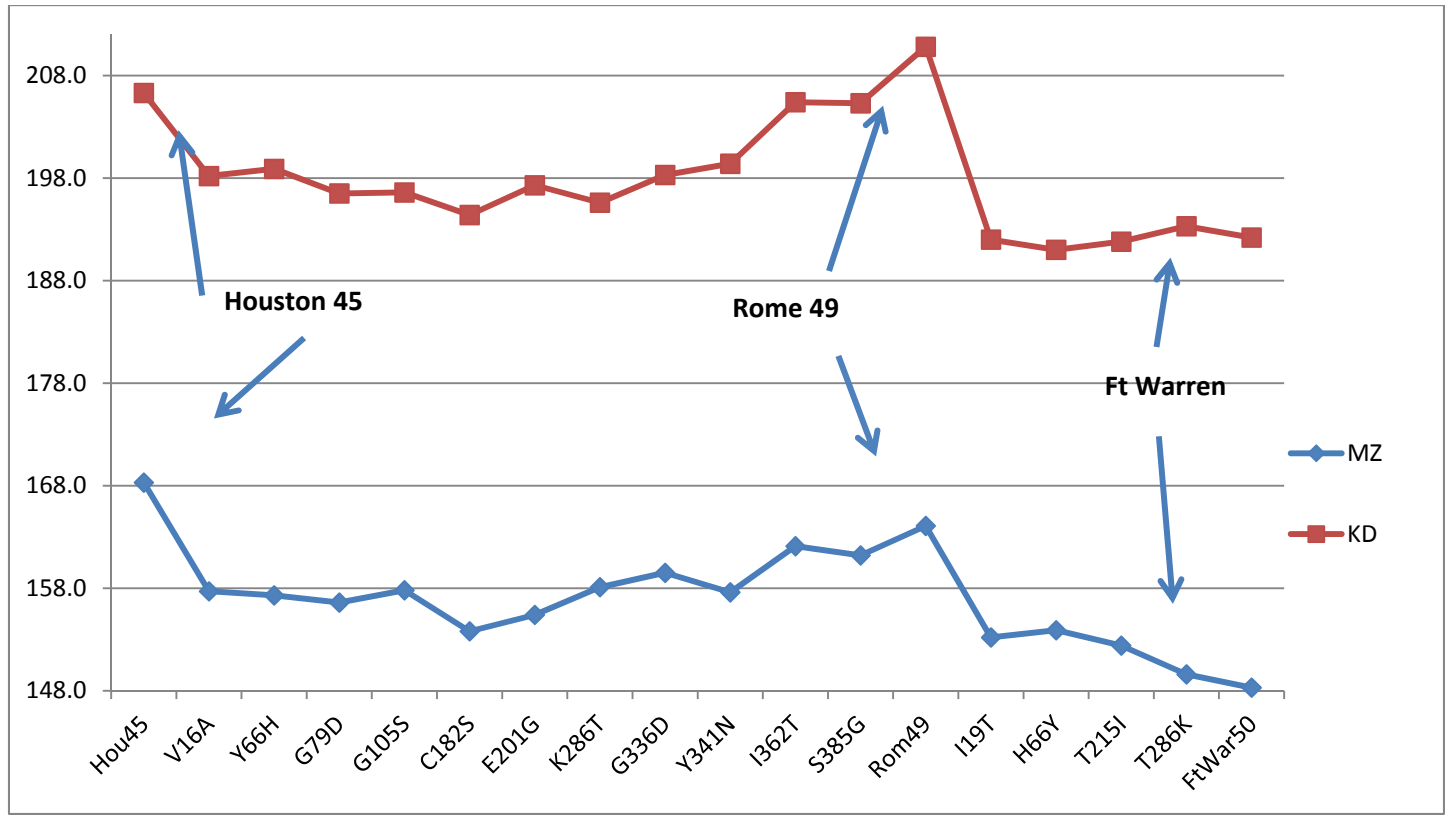

Fig. 3. A sketch shows the effects of individual mutations leading up to, and including, the first 1949-1950 A-B vaccination punctuation. The dominant effect of I19T, near the center of the 7-35 transmembrane segment, is clear.

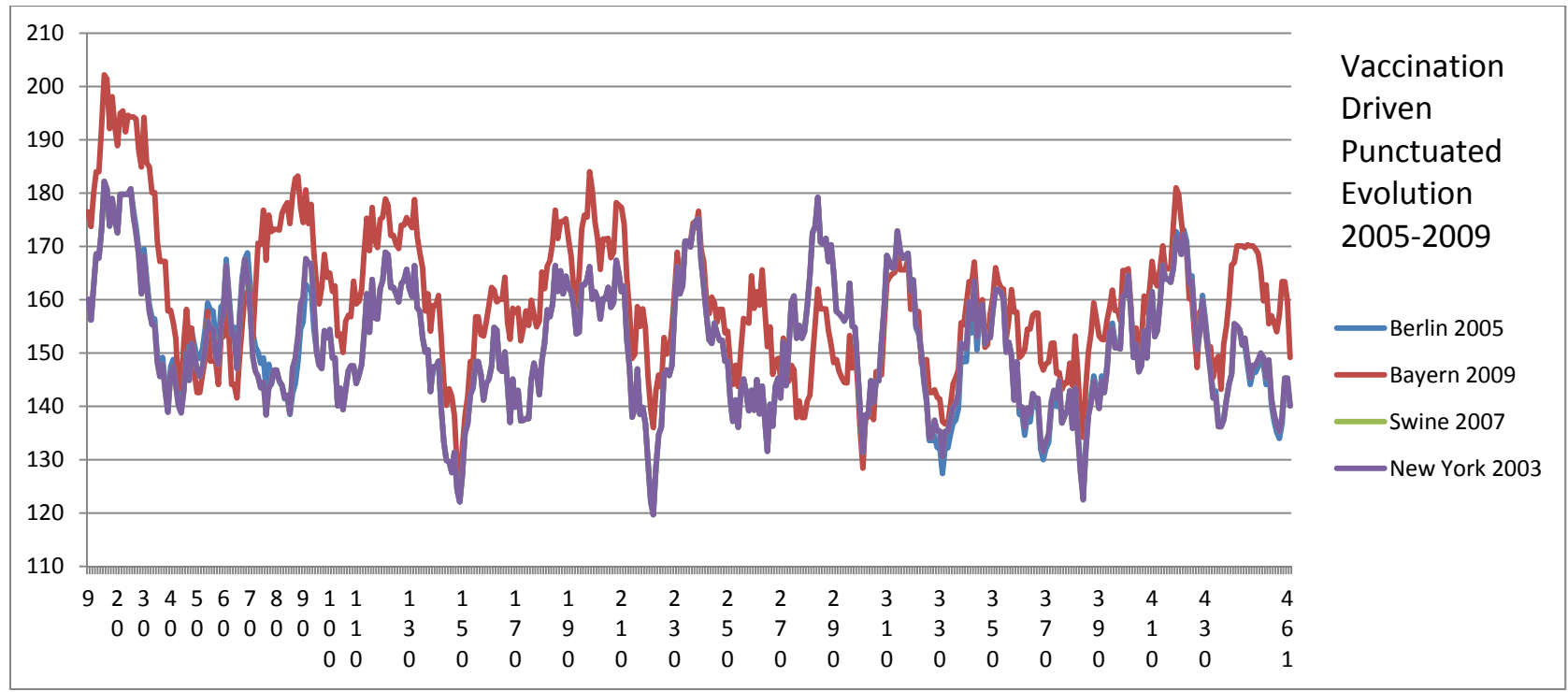

Fig. 4. In Fig. 1 the effects of the first A-B 1949-1950 vaccination-driven NA punctuation were confined to a few peaks and valleys of $<\psi(j)17>$, but the C-D 2003-2009 third vaccination-driven punctuation is qualitatively different and much more complex. It involves surprising post-vaccination increases in hydrophobic peaks (except in the central region 230-310), corresponding to overall globular compaction, yet there is still global smoothing, which continues the trend started by the A-B punctuation. Here the pre-vaccination chain profile differences are scarcely visible, but they can be resolved by calculating hydropathic roughness $R_{MZ}(17)$. The 2009 flattening between 210 and 300 could facilitate cleavage of sialic acid, by making its binding site in the center of HA more closely resemble a cutting board.

|  |  | Superstrains | | |
|---|---|---|---|---|
|  |  | MZ | KD |  |
| 18-45 | Aver | 163.8 | 211.3 | A |
|  | σ | 3.8 | 9.8 |  |
| 50-57 | Aver | 147.6 | 192.2 | B |
|  | σ | 1.7 | 4.5 |  |
| 76 (Fort Dix) | Aver | 169.7 | 236.0 | AR |
|  | σ |  |  |  |
| 86 | Aver | 164.7 | 214.0 | A′ |
|  | σ | 3.1 | 4.3 |  |
| 88 | Hokkaido | 148.7 | 201.4 | T |
|  | σ |  |  |  |
| 89-03 | Aver | 141.5 | 198.5 | C |
|  | σ | 1.9 | 4.4 |  |
| 05-07 | Aver | 145.5 | 197.0 | swine flu |
|  | σ | 2.5 | 7.4 |  |
| 09-10 | Aver | 124.8 | 174.7 | D |
|  | σ | 0.4 | 0.6 |  |

Table I.  Superstrain roughness $r_{MZ}$(17) plateaus and vaccination/migration transitions.  The time periods are abbreviated (1918-1945 is written as 18-45, and 2009 as 09).  The average values and standard deviations σ are estimated from small samples, but are thought to be accurate enough to exhibit the main features of both vaccination and migration, as described in the text.  Four superstrain plateaus (A-D) are evident.  Also indicated are the Fort Dix reversion, the transition T from A′ to C, and the incipient swine flu

pandemic.   The overall trend towards reducing roughness (MZ drops from ~ 164 to ~ 120 {D*, just emerging in Wash DC 2010-2011}) by large-scale vaccinations, which has overriden migration pressures, is clear.   A more detailed comparison of 20 strains (1918-2010) using the MZ and KD scales showed a 90% correlation.   However, the even more detailed Hawaii 2007 analysis in the text shows that the MZ scale is about 3x more accurate than the KD scale for NA.

amino acid; profile; hydropathic; virus; vaccination; evolution; punctuation; glycoprotein; sialic acid; neuraminidase; convergence; mutation prolific; global; engineer; oncolytic; network; self-organized; kinetics; N-glycan; oligomer; sliding window; variance;